\documentclass[preprint,preprintnumbers,amsmath,amssymb]{revtex4}

\usepackage{graphicx}
\usepackage{dcolumn}
\usepackage{bm}

\begin{document}


\title{Magnetocrystalline anisotropy and antiferromagnetic phase transition in PrRh$_{2}$Si$_{2}$}
\author{V K Anand}
\email{vivekkranand@gmail.com}
\author{Z Hossain}
\affiliation{Department of Physics, Indian Institute of Technology, Kanpur 208016, India}
\author{G Behr}
\affiliation{Institute for Solid State and Materials Research, Dresden, Germany}
\author{G Chen}
\author{M Nicklas}
\author{C Geibel}
\affiliation{Max Planck Institute for Chemical Physics of Solids, 01187 Dresden, Germany}
\date{\today}

\begin{abstract}

We present magnetic and transport properties of PrRh$_{2}$Si$_{2}$ single crystals which exhibit antiferromagnetic order below T$_{N}$ = 68 K. Well defined anomalies due to magnetic phase transition are observed in magnetic susceptibility, resistivity, and specific heat data. The T$_{N}$ of 68 K for PrRh$_{2}$Si$_{2}$ is much higher than 5.4 K expected on the basis of de-Gennes scaling. The magnetic susceptibility data reveal strong uniaxial anisotropy in this compound similar to that of PrCo$_{2}$Si$_{2}$. With increasing pressure T$_{N}$ increases monotonically up to T$_N$~=~71.5 K at 22.5 kbar.

\end{abstract}

\maketitle


\section*{Introduction}

YbRh$_{2}$Si$_{2}$ has been widely investigated due to its proximity to a quantum phase
transition \cite{1,2,3,4}. We show in this paper that its Pr-homolog PrRh$_{2}$Si$_{2}$ also
presents unique magnetic properties. All the investigated RRh$_{2}$Si$_{2}$ (R = rare
earth) compounds have been found to order antiferromagnetically \cite{1,5,6,7,8,9,10,11,12}. Among them
GdRh$_{2}$Si$_{2}$ has the highest ordering temperature, T$_{N}$ $\sim$ 106~K \cite{6}.
EuRh$_{2}$Si$_{2}$ exhibits complex magnetic order with an antiferromagnetic ordering below
25 K \cite{11}. CeRh$_{2}$Si$_{2}$ and YbRh$_{2}$Si$_{2}$ have unusual and interesting
magnetic properties which are discussed below.

The antiferromagnetic ordering temperature T$_{N }$ $\sim$ 36 K in CeRh$_{2}$Si$_{2}$ is
very high compared to the de-Gennes expected ordering temperature of 1.2 K \cite{12,13}.
One more transition is observed at 24 K. The exact nature (localized versus itinerant) of
the magnetism of CeRh$_{2}$Si$_{2}$ is not yet settled. The pressure dependence of T$_{N}$
and of the magnetic moment indicates an itinerant nature of the magnetism \cite{13}. The
itinerant character of magnetism in CeRh$_{2}$Si$_{2}$ has also been suggested from a
systematic study of doping at Rh sites in Ce(Rh$_{1-x}$Pd$_{x}$)$_{2}$Si$_{2}$ \cite{14}.
However, the dHvA study suggests local moment magnetism in CeRh$_{2}$Si$_{2}$ at ambient
pressure. Under the application of pressure the Fermi surface topology changes
discontinuously leading to an itinerant moment magnetism above the critical pressure of
around 1 GPa \cite{15}.  Pressure induced superconductivity has been observed around 1 GPa
below 0.5 K \cite{16,17}.

Heavy-fermion YbRh$_{2}$Si$_{2}$ has an antiferromagnetic ordering
temperature T$_{N}$ of $\sim$ 70 mK \cite{1}. The antiferromagnetic order can be suppressed
very easily by application of magnetic field or by substitution of Si by Ge, leading to a
quantum critical point \cite{2,3,4}. Electrical transport, thermodynamic and thermal expansion data reveal that quantum critical point in YbRh$_{2}$Si$_{2}$ is of local nature in contrast to the spin density wave type quantum critical point in CeCu$_{2}$Si$_{2}$ \cite{18,19}.

Crystal field effects can have strong influence on the properties of Pr-compounds. For
example, the low lying crystal field excitations are responsible for the heavy fermion
behavior in unconventional superconductor PrOs$_{4}$Sb$_{12}$ \cite{20,21,22}.  Despite numerous
investigations on RRh$_{2}$Si$_{2}$, we did not find any discussion in literature on the
properties of PrRh$_{2}$Si$_{2}$. In this paper we report magnetization, specific heat,
electrical resistivity and magnetoresistance of PrRh$_{2}$Si$_{2}$ single crystals. In
addition, we also carried out pressure dependent electrical resistivity measurements.

\section*{Sample preparation and measurements}

Single crystals of PrRh$_{2}$Si$_{2}$ were grown from indium flux as well as using
floating zone method in a mirror furnace (CSI Japan). Appropriate amounts of high purity elements (Pr: 99.99\%, La: 99.9\%, Rh: 99.999\% and Si: 99.9999\%) were arc melted several times on a water cooled copper hearth under argon atmosphere. The arc melted polycrystalline
PrRh$_{2}$Si$_{2}$ and indium were taken in a molar ratio of 1:20 in an alumina crucible,
which was then sealed inside a tantalum crucible with a partial pressure of argon gas. The
sealed tantalum crucible was heated to 1450 $^{o}$C under argon atmosphere for two hours
and then cooled down to 900 $^{o}$C at a rate of 5 $^{o}$C/hour. Below 900$^{o}$C rate of
cooling was increased to 300 $^{o}$C/hour. Indium flux was removed by etching with dilute
hydrochloric acid. We obtained single crystals of about 2.5 mm x 1.5 mm x 0.4 mm by this
method. We also succeeded in growing PrRh$_{2}$Si$_{2}$ single crystal using float zone
mirror furnace using 10 mm/h growth rate and counter-rotation of seed and feed rods.
The diameter of the float zone grown crystal was about 6 mm.

Samples were characterized by copper K$_\alpha$ X-ray diffraction and scanning electron microscope (SEM) equipped with energy dispersive X-ray analysis (EDAX). Laue method was used to orient the single crystals. A commercial SQUID magnetometer was used to measure magnetization. Specific heat was measured using relaxation method in a physical property measurement system (PPMS--Quantum design). Electrical resistivity was measured by standard ac four probe technique using AC-transport option of PPMS. Pressure studies of the electrical resistivity up to 2.3 GPa and in the temperature range 3 K $<$ T $<$ 300 K were carried out utilizing a clamp-type double layer pressure cell consisting of an inner cylinder made of NiCrAl and an outer body of Cu:Be. Silicone oil served as pressure
transmitting medium. The pressure inside the cell was determined at low temperature by the
inductively measured shift of the superconducting transition temperature of lead.

\section*{Results and Discussion}

From the analysis of powder X-ray diffraction data of the crushed single crystals (figure 1), we find that PrRh$_{2}$Si$_{2}$ crystallizes in ThCr$_{2}$Si$_{2}$-type tetragonal
structure (space group I4/mmm) with the lattice parameters \textit{a} = 0.4079 nm,
\textit{c} = 1.0138 nm, and the unit cell volume = 0.16876 nm$^{3}$ for the
flux grown sample, and \textit{a} = 0.4078 nm, \textit{c} = 1.0138 nm, and the
unit cell volume = 0.16858 nm$^{3}$ for the float zone grown sample. The X-ray diffraction and SEM image confirmed the samples to be single phase. The EDAX composition analysis confirmed the desired stoichiometric composition of 1:2:2.

\begin{figure}
\centering
\includegraphics[width=14cm, keepaspectratio]{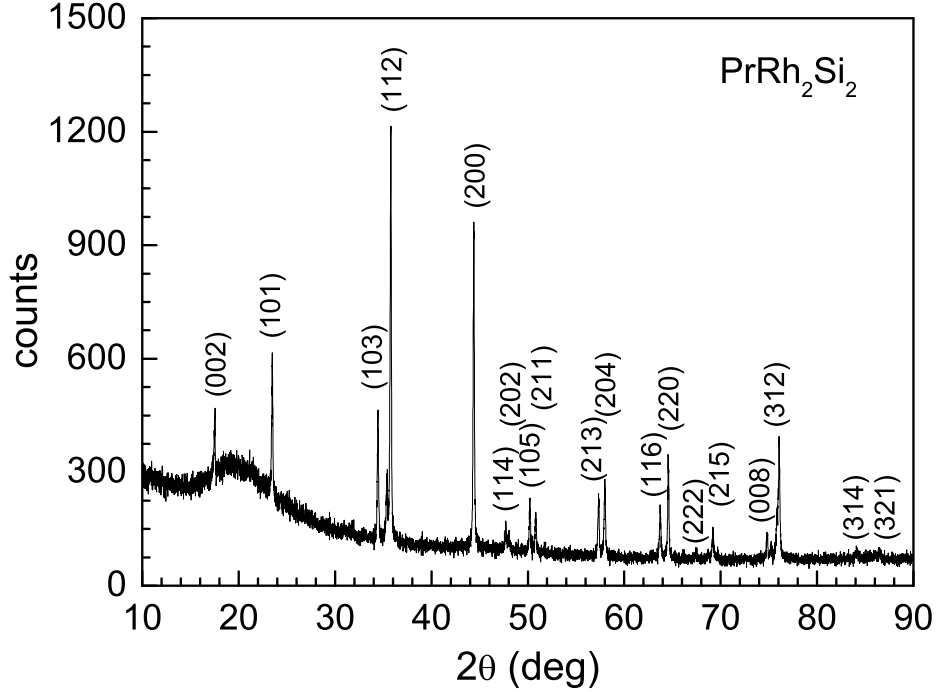}
\caption{\label{fig1} Indexed powder x-ray diffraction pattern of ThCr$_{2}$Si$_{2}$-type
body-centered-tetragonal pulverized PrRh$_{2}$Si$_{2}$ single crystal (flux grown).}
\end{figure}

\begin{figure}
\centering
\includegraphics[width=12cm, keepaspectratio]{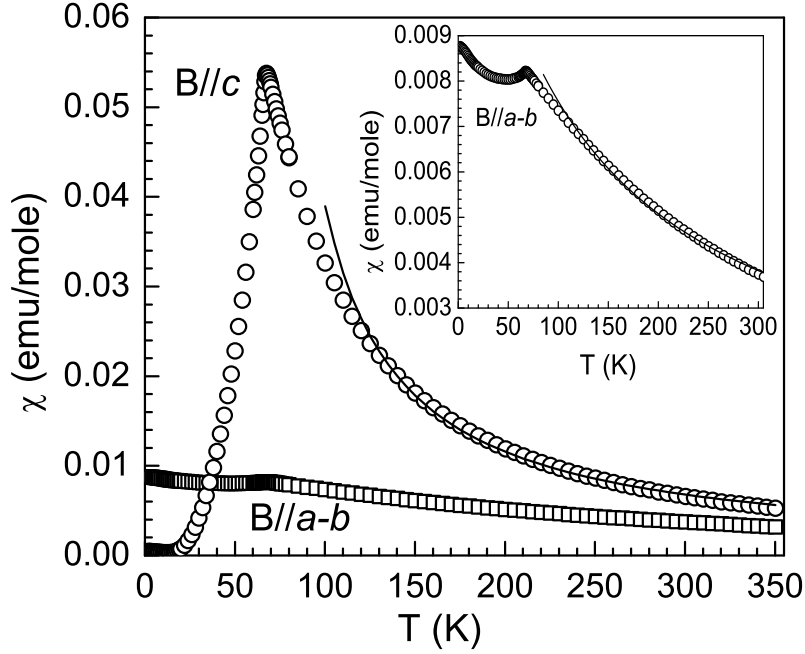}
\caption{\label{fig2} Temperature dependence of magnetic susceptibility of
PrRh$_{2}$Si$_{2}$ single crystal (flux grown) measured in a field of 3.0 T. Inset shows
the enlarged view of B//\textit{a-b} data. The solid lines represent fit to Curie-Weiss behaviour.}
\end{figure}

The temperature dependence of the magnetic susceptibility of PrRh$_{2}$Si$_{2}$
single crystal is shown in figure 2 for magnetic field applied along the
\textit{a-b} plane and the \textit{c}-axis. A large anisotropy in the magnetic
susceptibility $\chi$(T) is observed. The susceptibility data have much larger
values for  B//\textit{c} compared to that for B//\textit{a-b} implying the easy
axis to be the \textit{c}-axis. This anisotropic behavior is similar to the
strong uniaxial anisotropy along \textit{c}-axis in CeRh$_{2}$Si$_{2 }$ \cite{10}
but different from the easy plane behavior observed in YbRh$_{2}$Si$_{2}$
\cite{1}. Within a series of R-T-X compound, the change of the magnetic
anisotropy with changing the R-element is governed by the change in
$\alpha_{J }$ second order Stevens factor within the CEF Hamiltonian \cite{23}.
A change from a uniaxial behavior in CeRh$_{2}$Si$_{2}$ and PrRh$_{2}$Si$_{2}$ to
an easy plane behavior in YbRh$_{2}$Si$_{2 }$ is in full accordance
with $\alpha_{J}$ $<$ 0 for Ce- and Pr- while $\alpha_{J}$ $>$ 0 for Yb-compound.
Since in all three cases the anisotropy is very pronounced, it indicates a very large and positive A$_{2}^{0}$ CEF-parameter in the whole RRh$_{2}$Si$_{2}$ series. An
antiferromagnetic transition is observed in the susceptibility data at 68 K for both
B//\textit{a-b} and B//\textit{c}. As expected for an antiferromagnet T$_N$ decreases with increasing
magnetic field (T$_N$ = 66.5 K at B = 5 T). The susceptibility data exhibit slight deviation from the Curie-Weiss behaviour $\chi$(T) = C/(T-$\theta_{p}$) for both B//\textit{a-b} and B//\textit{c} due to the effect of crystal fields. From the linear fit of inverse susceptibility data (100 K -- 300 K) at 3 T we obtain the effective magnetic moment $\mu_{eff}$ = 3.48 $\mu_{B}$ (very close to the theoretical value of 3.58 $\mu_{B}$ for Pr$^{3+}$ ions) and the Curie-Weiss temperature $\theta_{p}^{a}$ = -103.2 K for B//\textit{a-b}, and $\mu_{eff}$ = 3.63 $\mu_{B}$ and $\theta_{p}^{c}$ = +57.9 K for B//\textit{c}. Further, we note a very pronounced peak and a rapid decrease of magnetic susceptibility to essentially zero value below 20 K for B//\textit{c} and a much weaker temperature dependence for B//\textit{a-b}, which suggests strongly anisotropic Ising-type antiferromagnetism in PrRh$_{2}$Si$_{2}$ similar to that of PrCo$_{2}$Si$_{2}$ \cite{24}.

The isothermal magnetization data exhibit a linear dependence on field at 60 K
(magnetically ordered state) and 80 K (paramagnetic state) for both B//\textit{a-b} and
B//\textit{c} (figure 3). The magnetic moments at 5 T are very small in both the
directions (0.07 $\mu_{B}$/Pr for B//\textit{a-b} and 0.39 $\mu_{B}$/Pr for B//\textit{c}) and the maximum value attained is only 12\% of the saturation magnetization for Pr$^{3+}$ ion (3.2 $\mu_{B}$/Pr). Measurements at higher fields are required to observe the metamagnetic transitions which are expected in antiferromagnets with strong magneto-crystalline anisotropy.

\begin{figure}
\centering
\includegraphics[width=12cm, keepaspectratio]{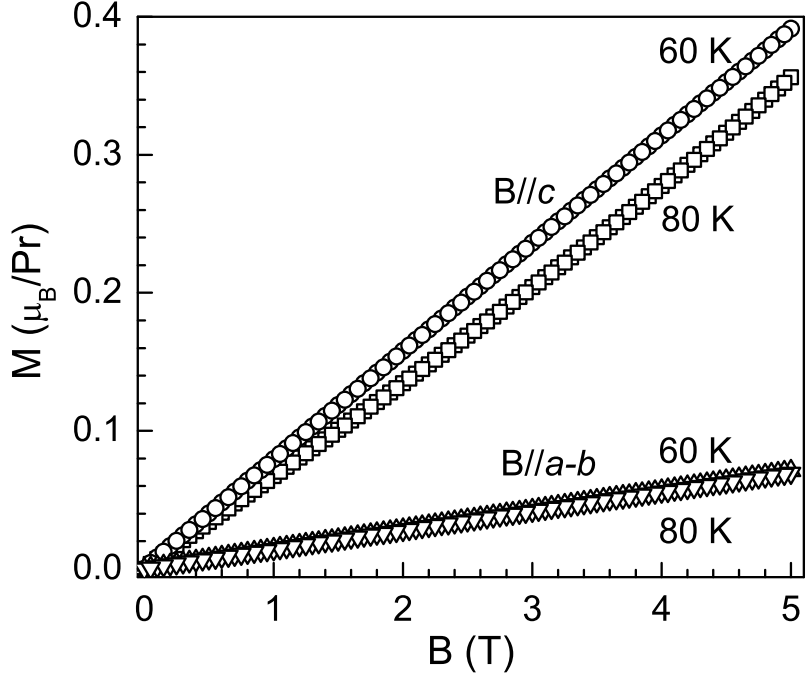}
\caption{\label{fig3} Field dependence of isothermal magnetization of PrRh$_{2}$Si$_{2}$
single crystal (flux grown) at 60 and 80 K along B//\textit{c} and B//\textit{a-b}.}
\end{figure}

\begin{figure}
\centering
\includegraphics[width=12cm, keepaspectratio]{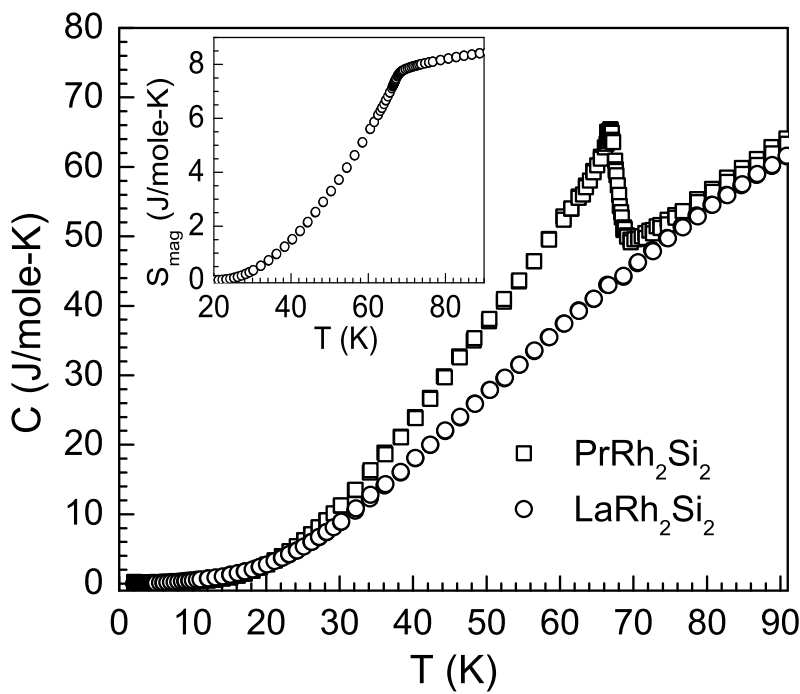}
\caption{\label{fig4} Temperature dependence of the specific heat of single crystal
PrRh$_{2}$Si$_{2}$ (flux grown) and polycrystalline LaRh$_{2}$Si$_{2}$ in the temperature
range 2 to 90 K. The inset shows the magnetic contribution to the entropy of
PrRh$_{2}$Si$_{2}$.}
\end{figure}

The specific heat data of single crystal PrRh$_{2}$Si$_{2}$ (indium flux grown) together
with that of the nonmagnetic reference compound LaRh$_{2}$Si$_{2}$ are shown in figure 4. The specific heat of PrRh$_{2}$Si$_{2}$ exhibits a pronounced $\lambda$-type anomaly at 68 K, which confirms the intrinsic nature of antiferromagnetic ordering in this compound. The
float zone grown single crystal of PrRh$_{2}$Si$_{2}$ also exhibits a similar well defined
anomaly at 68 K due to antiferromagnetic order. The specific heat data of PrRh$_{2}$Si$_{2}$ and LaRh$_{2}$Si$_{2}$ hardly differ from each other below 20 K showing
that  the magnetic excitations have vanished exponentially below 20 K. This indicates a
large gap in the magnetic excitation spectra in the ordered state, which can obviously be
attributed to the strong Ising-type anisotropy observed in the magnetic susceptibility
data. The linear coefficient to the specific heat is $\gamma$ $\sim$ 18 mJ/mole-K$^{2}$.
The temperature dependence of the magnetic entropy is shown as inset in figure 4.
At 70 K the magnetic entropy attains a value of 7.85 J/mole-K, which is 36\% more
than $R$ln2 and 14 \% lower than $R$ln3. Thus, either three singlets or one singlet and one doublet CEF levels are in the energy-range below 80 K and involved in the
magnetic ordering. Because of the huge uniaxial anisotropy and the general trend of the CEF parameters within the RRh$_{2}$Si$_{2}$ series, one can suspect that these lowest CEF levels are the two $\Gamma_{1}$ singlets and either the $\Gamma_{2}$ singlet or the $\Gamma_{5}$ doublet \cite{25}.

\begin{figure}
\centering
\includegraphics[width=12cm, keepaspectratio]{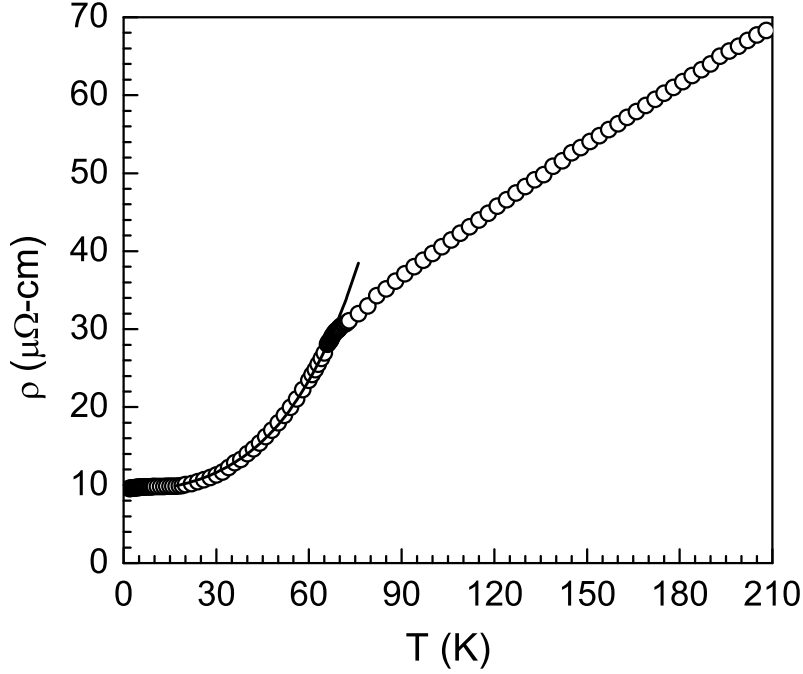}
\caption{\label{fig5} Temperature dependence of electrical resistivity (I//\textit{a-b})of
flux grown PrRh$_{2}$Si$_{2}$ single crystal in the temperature range 1.8 -- 210 K.
Solid line shows the fit to gapped magnon characteristics in the ordered
state, i.e. $\rho(T) = \rho_{0} + A T^2 + C \bigg\{ \frac{1}{5} T^5 + \triangle  T^4 + \frac{5}{3}
\triangle ^2 T^3 \bigg\} exp(-\triangle /T)$.}
\end{figure}

The electrical resistivity measured with ac current flowing in the \textit{a-b}
plane is shown in figure 5. The resistivity shows typical metallic behavior with
room temperature resistivity $\rho_{300 K }$ of 85 $\mu\Omega$-cm, residual
resistivity $\rho_{0}$ $\sim$ 9.6 $\mu\Omega$-cm and residual resistivity ratio
(RRR) $\sim$ 9. A linear decrease of resistivity is observed with decreasing
temperature until it meets the antiferromagnetic transition at 68 K, below which
the resistivity shows a large decrease. In the ordered state the resistivity data
present gapped magnon characteristics and fit well to the relation \cite{26}

\begin{displaymath}
\rho(T) = \rho_{0} + A T^2 + C \bigg\{ \frac{1}{5} T^5 + \Delta  T^4 + \frac{5}{3}
\Delta ^2 T^3 \bigg\} exp(-\Delta /T)
\end{displaymath}

below 65 K (inset of figure 5) where $\rho_{0 }$ = 9.8 $\mu\Omega$-cm is the
residual resistivity, A = 0.00241 $\mu\Omega$-cm/K$^{2}$ is the coefficient to the
Fermi liquid term, C = 8.96 x 10$^{-9}$ $\mu\Omega$-cm/K$^{5}$ is the prefactor to the magnon contribution, and $\Delta$ = 37.8 K is the magnon energy gap.

As both CeRh$_{2}$Si$_{2}$ and YbRh$_{2}$Si$_{2}$ exhibit strong pressure dependence in
the electrical resistivity we have performed resistivity measurement on
PrRh$_{2}$Si$_{2}$ under externally applied pressure.  Up to 22.5 kbar there is no pronounced effect of externally applied pressure on the resistivity except an increase of T$_{N}$ from 68.5 K at p = 0 to 71.5 K at p = 22.5 kbar (figure 6). Similar weak effect of pressure on the magnetically ordered state was also found in PrCo$_{2}$Si$_{2}$ \cite{27}.

\begin{figure}
\centering
\includegraphics[width=12cm, keepaspectratio]{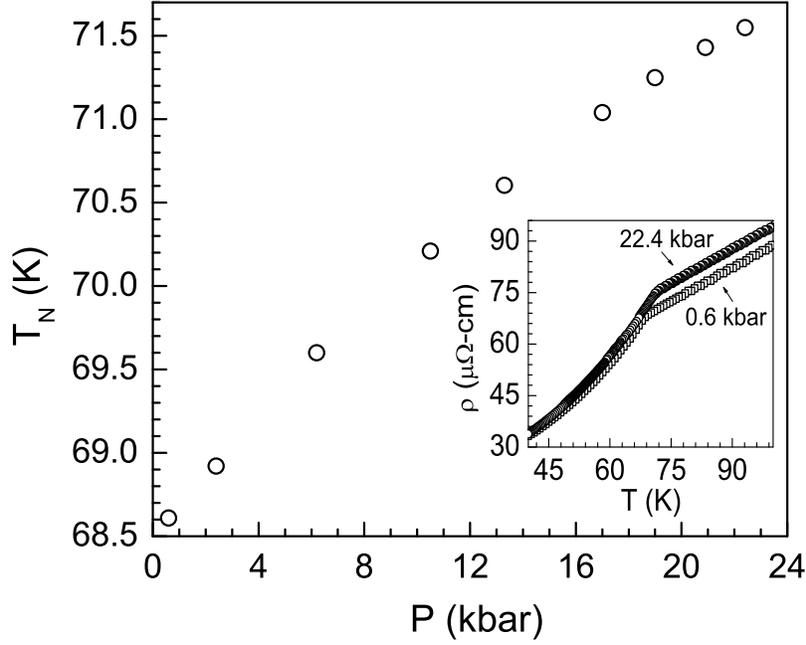}
\caption{\label{fig6} T$_{N}$ of PrRh$_{2}$Si$_{2}$ as a function of externally applied
pressure. The inset shows temperature dependence of resistivity under pressure.}
\end{figure}

From the de-Gennes scaling in the family of RRh$_{2}$Si$_{2}$ (R = rare earths) one would
expect an ordering temperature of 5.4 K in PrRh$_{2}$Si$_{2}$. While in CeRh$_{2}$Si$_{2}$
the anomalously high T$_{N}$ might be a result of the mixture of localized and itinerant
character of the magnetic order we can not offer any clear reason for the high T$_{N}$ of
PrRh$_{2}$Si$_{2}$. Enhanced density of states as in the case of GdRh$_{2}$Si$_{2}$ and
large value of exchange constant (as evidenced by large $\theta_{p}$) definitely contribute
to higher value of T$_{N}$. It is also found that RRh$_{2}$Si$_{2}$ compounds which have
higher values of T$_{N}$ than expected on the basis of de-Gennes scaling have their moments
aligned along \textit{c}-axis below T$_{N}$. PrRh$_{2}$Si$_{2}$ also has higher T$_{N}$
than expected and the magnetic susceptibility data suggest that Pr moments lie along
\textit{c}-axis in this case also. We suspect the uniaxial anisotropy which forces the
moment to
lie along the \textit{c}-axis is also responsible for the high T$_{N}$ in PrRh$_{2}$Si$_{2}$.
System with uniaxial anisotropy has much larger value of magnetic susceptibility for
B//(easy-axis) which helps in the process of magnetic ordering. Thus T$_{N}$ for a system
with uniaxial anisotropy will be higher than that of an isotropic system or a weakly
anisotropic system.

\section*{Conclusion}

We succeeded in growing single crystals of PrRh$_{2}$Si$_{2}$ which forms in
ThCr$_{2}$Si$_{2}$-type body- centered tetragonal structure. Temperature dependent magnetic
susceptibility, electrical resistivity, specific heat data reveal strongly anisotropic
Ising type antiferromagnetic order below 68 K in this compound. Application of pressure up
to 22.5 kbar does not stabilize any new ordered phase but T$_{N}$ increases from 68 K to
71.5 K.

\section*{Acknowledgement}

Technical assistance from Mr. Jochen Werner is gratefully acknowledged.


\begin{thebibliography}{27}

\bibitem {1}
  Trovarelli O, Geibel C, Mederle S, Langhammer C, Grosche F M, Gegenwart P, Lang M, Sparn
  G and Steglich F 2000 {\it Phys. Rev. Lett.} {\bf 85} 626

\bibitem {2}
  Gegenwart P, Custers J, Geibel C, Neumaier K, Tayama T, Tenya K, Trovarelli O and
  Steglich F 2002 {\it Phys. Rev. Lett.} {\bf 89} 056402

\bibitem {3}
  Plessel J, Abd-Elmeguid M M, Sanchez J P, Knebel G, Geibel C, Trovarelli O and Steglich F
  2003 {\it Phys. Rev. B} {\bf 67} 180403(R)

\bibitem {4}
  Custers J, Gegenwart P, Wilhelm H, Neumaier K, Tokiwa Y, Trovarelli O, Geibel C, Steglich
  F, P\'{e}pln C and Coleman P 2003 {\it Nature} {\bf 424} 524

\bibitem {5}
 Felner I and Nowik I 1983 {\it Solid State Commun.} {\bf 47} 831

\bibitem {6}
 Tung L D, Franse J J M, Buschow K H J, Brommer P E and Thuy N P 1997 {\it J. Alloys Comp}
 {\bf 260} 35

\bibitem {7}
 Szytula A, \`{S}laski M, Ptasiewicz-Bak H, Leciejewicz J and Zygmunt A 1984 {\it Solid
 State Commun.} {\bf 52} 395

\bibitem {8}
 \`{S}laski M, Leciejewicz J and Szytula A 1983 {\it J. Magn. Magn. Mater.} {\bf 39} 268

\bibitem {9}
 Melamud M, Pinto H, Felner I and Shaked H 1984 {\it J. Appl. Phys.} {\bf 55} 2034.

\bibitem {10}
 Quezel S, Rossat-Mignod J, Chevalier B, Lejay P and Etourneau J 1984 {\it Solid State
 Commun.} {\bf 49} 685

\bibitem {11}
 Hossain Z, Trovarelli O, Geibel C and Steglich F 2001 {\it J. Alloys Comp.} {\bf 323-324}
 396

\bibitem {12}
 Graf T, Hundley M F, Modler R, Movshovich R, Thompson J D, Mandrus D, Fisher R A and
 Phillips N E 1998 {\it Phys. Rev. B} {\bf 57} 7442

\bibitem {13}
 Kawarazaki S, Sato M, Miyako Y, Chigusa N, Watanabe K, Metoki N, Koike Y and Nishi M 2000
 {\it Phys. Rev. B} {\bf 61} 4167

\bibitem {14}
 G\'{o}mez Berisso M, Pedrazzini P, Sereni J G, Trovarelli O, Geibel C and Steglich F 2002
 {\it Eur. Phys. J. B} {\bf 30} 343

\bibitem {15}
 Araki S, Settai R, Kobayashi T C, Harima H and \={O}nuki Y 2001 {\it Phys. Rev. B} {\bf
 64} 224417

\bibitem {16}
 Movshovich R, Graf T, Mandrus D, Thompson J D, Smith J L and Fisk Z 1996 {\it Phys. Rev.
 B} {\bf 53} 8241

\bibitem {17}
 Araki S, Nakashima M, Settai R, Kobayashi T C and \={O}nuki Y 2002 {\it J. Phys.: Condens.
 Matter}  {\bf 14} L377

\bibitem {18}
 Gegenwart P, Langhammer C, Geibel C, Helfrich R, Lang M, Sparn G, Steglich F, Horn R,
 Donnevert L, Link A and Assmus W 1998 {\it Phys. Rev. Lett.} {\bf 81} 1501

\bibitem {19}
  Stockert O,  Faulhaber E, Zwicknagl G, St\"{u}{\ss}er N, Jeevan H S, Deppe M, Borth R,
  K\"{u}chler R, Loewenhaupt M, Geibel C and Steglich F 2004 {\it Phys. Rev. Lett.} {\bf
  92} 136401

\bibitem {20}
 Bauer E D, Frederick N A, Ho P -C, Zapf V S and Maple M B 2002 {\it Phys. Rev. B} {\bf 65}
 100506(R)

\bibitem {21}
 Goremychkin E A, Osborn R, Bauer E D, Maple M B, Frederick N A, Yuhasz W M,
Woodward F M, and Lynn J W 2004 {\it Phys. Rev. Lett.} {\bf 93} 157003

\bibitem {22}
 Thalmeier P 2006 {\it Physica B} {\bf 378-380} 261

\bibitem{23}
 see e.g. Buschow K H J and de Boer F R 2003 {\it Physics of Magnetism and Magnetic
 Materials} (Kluwer academic/Plenum publishers) p~56

\bibitem {24}
 Shigeoka T, Fujii H, Yonenobu K, Sugiyama K and Date M 1989 {\it J. Phys. Soc. Japan} {\bf
 58} 394

\bibitem{25}
 Santini P, Amoretti G, Blaise A and Caciuffo R 1993 {\it J. Appl. Phys.} {\bf 73}  6560

\bibitem {26}
 Jobiliong E, Brooks J S, Choi E S, Lee H and Fisk Z 2005 {\it Phys. Rev. B} {\bf 72}
 104428

\bibitem {27}
 Kawano S, Onodera A, Achiwa N, Nakai Y, Shigeoka T and Iwata N 1995 {\it Physica B} {\bf
 213-214} 321

\end{thebibliography}
\end{document}